\documentclass[fleqn,10pt]{wlscirep} 
\newcommand{\deriv}[2]{\frac{\partial^{#2}}{\partial{#1}^{#2}}} 
\usepackage{nicefrac}

\title{Two-fluid, hydrodynamic model for spherical electrolyte systems}

\author[1,*]{Christin David}
\affil[1]{Madrid Institute for Advanced Studies in Nanoscience (IMDEA Nanoscience), C/ Faraday 9, 28049 Madrid, Spain}
\affil[*]{christin.david@imdea.org}

\begin{abstract}
Spatial interaction effects between charge carriers in ionic systems play a sizable role beyond a classical Maxwellian description. We develop a nonlocal, two-fluid, hydrodynamic theory of charges and study ionic plasmon effects, i.~e. collective charge oscillations in electrolytes. Ionic spatial dispersion arises from both positive and negative charge dynamics with an impact in the (far-)infrared. Despite highly classical parameters, nonlocal quenching of up to 90\% is observed for particle sizes spanning orders of magnitude. Notably, the ionic system is widely tunable via ion concentration, mass and charge, in contrast to solid metal nanoparticles. A nonlocal soft plasmonic theory for ions is relevant for biological and chemical systems bridging hard and soft matter theory and allowing the investigation of non-classical effects in electrolytes in full analogy to solid metal particles. The presented semi-classical approach allows studying plasmonic photo-catalysis introducing nonlocal aspects into electrolyte-metal interactions.
\end{abstract}
\begin{document}

\flushbottom
\maketitle

\thispagestyle{empty}

\section*{Introduction}
The field of plasmonics is widely understood as the investigation of collective oscillations of the electron plasma in metallic systems induced by an external electromagnetic field. In a broader sense, such plasma oscillations can be found for any system where charged carriers are induced, such as stretchable~\cite{Zhang2014} and amorphous materials~\cite{Caputo2013} and charged plasmas in space~\cite{Peratt1992, Choudhuri1998}, e.~g. the high energy nuclear plasma in the solar core, the ion gas in the ionosphere or intergalactic clouds.

Electrolytes consist of positively and negatively charged ions even in an equilibrium situation. Hence, the question arises, if plasmon-like behavior can be observed in ionic systems and if interaction effects between the charge carriers play a sizable role as compared to the quantum effects observed for metal nanoparticles beyond a classical Maxwellian description~\cite{Scholl2012, Cirac2012, Savage2012, Raza2013, Raza2015b}.
Plasmons of noble metals are studied with respect to applications in sensing and spectroscopy~\cite{Gordon2008}, optical filters~\cite{Genet2007}, plasmonic colors~\cite{Kristensen2016}, lasers~\cite{Yu2008}, and quantum plasmonics~\cite{Bozhevolnyi2017a} exploiting the extreme confinement and sensitivity of their electro-optical excitation. The plasmonic properties of ions in solution, on the other hand, are of interest in biological and chemical systems. Local energy transfer and communication are central in the signaling and conductivity of nerve cells (axons)~\cite{Jacak2015} with great importance to the functionality of the biological system in question. Another active area potentially benefiting from this study is catalysis where the interaction of an electrolyte with its environment and in particular with functionalized surfaces is of central interest.

Next to classical nanophotonics used for ionic systems, we are in particular interested in non-classical effects. Soft plasmonics for spherical ions~\cite{Jacak2016a} and chains of axons~\cite{Jacak2015} have been studied within the RPA (Random Phase Approximation) method including Lorentz friction, i.~e., electron radiation due to the acceleration of charges in the oscillations.
Spatial dispersion phenomena such as charge (Coulomb) interaction and diffusion arise within the hydrodynamic model, where free charges are described within a linearized Navier-Stokes equation. This was extensively studied for metals~\cite{Eguiluz1975,Sipe1980,Ruppin1981,Fuchs1981,Fuchs1987,Rojas1988,Ruppin1989,Ruppin1992,Kreibig1995,Ruppin2001,Aizpurua2008,McMahon2010a,Raza2011,David2011a,Wiener2012,Ceglia2013a,David2013a,Luo2013,David2014a,Christensen2014,Mortensen2014,David2016a}, see for instance the review Ref.~\cite{Raza2015a}, and more recently in semiconductors~\cite{Maack2017}.
Core parameters are taken from ab initio methods such as DFT (Density Functional Theory)~\cite{Pines1952, Lang1970}, namely, the charge density distribution $n$ and the pressure $p$ between charges, which defines the nonlocal interaction strength $\beta$. 

\begin{figure}[ht!]
\centering
	\includegraphics[width=0.5\linewidth]{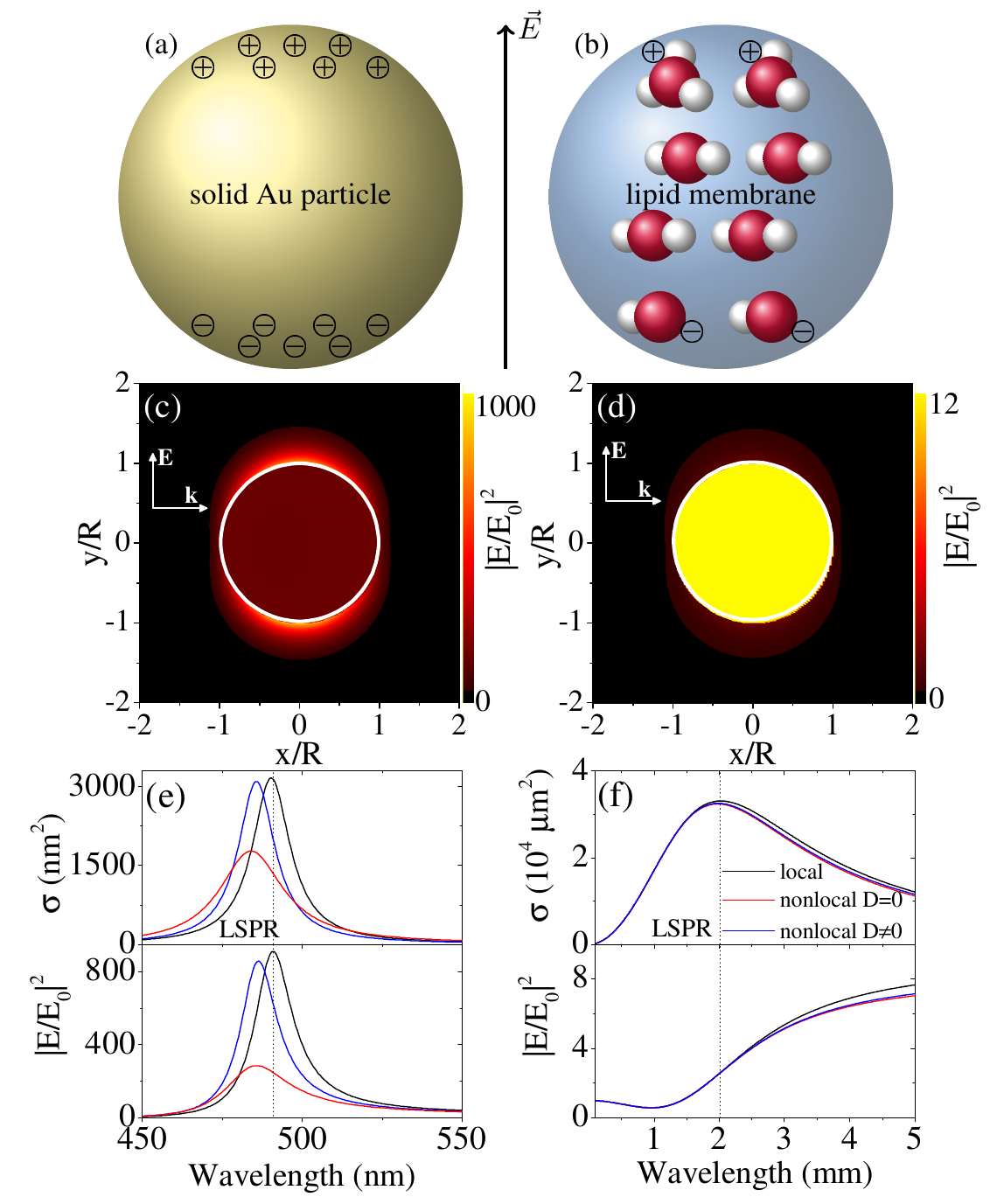}
    \caption{\textbf{Concept of (localized surface) plasmon resonance (LSPR) in ionic systems compared to metals.} Illustration of charge carrier displacement in (a) a metal solid and (b) an ionic fluid confined by e.~g. a lipid cellular membrane. (c), (d) Field enhancement map of the plasmonic resonance in both systems. (e), (f) Impact of nonlocal charge dynamics (with and without electron diffusion $D$) on metals and ions in terms of the extinction cross section (upper panel) and the field en\-hance\-ment (lower panel). The particle size is $R=10\ $nm and a Drude model for gold was used for clarity. The charge density is $n=n_+=n_-=1\nicefrac{\rm{mol}}{\rm{m}^3}N_A$.}
    \label{fig1}
\end{figure}

In an electrolyte, the charge carriers are heavy ions and both charge plasmas need to be considered. Such a two-fluid model of different charge carriers has been adapted to semiconductors recently~\cite{Maack2018}. Here, electrons and holes are described and their (effective) mass comprise an additional degree of freedom not available in metal systems. There is an imbalance of masses as for electrons and holes in semiconductors, but the masses of ions can be many orders of magnitude larger than the electron mass $m_e$. Ion concentration can be used to further tune the system similar to doping in semiconductor systems. In addition, depending on the choice of material, the charge $Q$ itself can be different from the elementary charge $e$. This adds yet another degree of freedom in the case of electrolytes. These recent findings offer interesting novel platforms to study nonlocal effects experimentally.
A plasmonic theory for ions in solution can bridge hard and soft matter theory and allow studying these interaction effects from a photonic perspective in full analogy to solid metal particles, see Fig.~\ref{fig1}. The semi-classical approach presented here can be fully integrated into standard nano-optic simulation frameworks and is considered to be of great interest for plasmonic photo-catalysis~\cite{Akimov2013} introducing nonlocal aspects into electrolyte-metal interactions possibly avoiding the hardship of DFT calculations.

In this article, we investigate ionic carriers in a finite electrolyte system made of hydronium H$_3$O$^+$ (more abundant than H$^+$) and hydroxide OH$^-$ at room temperature. Together with their masses and charges, this yields the parameters collected in table~\ref{tab.mat}. These can be created from confined ions within insulating membranes typically in the micrometer scale. In particular, the large ion mass and low concentration results in an energy and size regime very distinct from metal nanoparticles. However, we find that these systems, too, are highly tunable with various system parameters, such as the ion charge and mass ratio. Moreover, a remarkable impact of nonlocal interaction of charge carriers is found, resulting in particular in a reduction of the classically predicted intensity. Notably, this nonlocal quenching can be pushed to larger particle sizes by decreasing the ionic concentration, i.e., shifting the plasmon resonance frequency towards lower frequencies.

\section*{Results}
Plasmons in metals are an excitation of the conduction band electrons via a polarizing electromagnetic field that displaces the freely moving electrons with respect to the positively charged atomic cores, see Fig.~\ref{fig1}(a). In an electrolyte, we find even at thermal equilibrium a small density $n_\pm$ of ions of different charge $Q_\pm$ and mass $m_\pm$, see Fig.~\ref{fig1}(b). However, in the absence of a rigid crystal lattice, both types of ions can oscillate with typically different plasmon frequencies $\omega_{p\pm}^2=\nicefrac{4\pi Q_\pm^2 n_\pm}{m_\pm}$. The coupled system yields an effective localized surface plasmon resonance (LSPR) for spherical geometries at $\omega_{LSPR} = \sqrt{\nicefrac{(\omega_{p+}^2+\omega_{p-}^2)}{3\epsilon_b}}$, where $\epsilon_b$ is the background permittivity inside the confining membrane. In contrast to metals, where the mass equals the electron mass $m=m_e$ and the charge is the elementary charge $Q=e$, such ion systems become highly tunable through the choice of materials and their concentration. Assuming self-ionization of water at the thermodynamic equilibrium, we arrive at a concentration $n_{ThermEq}=6.022\times10^{16}$m$^{-3}$ for each type of ion. The localized plasmon resonance for this low density hydroxide and hydronium system, see table~\ref{tab.mat} for further parameters, results in $\lambda_{LSPR}=6.38$m ($\omega_{LSPR}=4.7\times10^7$Hz).

The plasmons of metals are of great interest due to their ability to capture electromagnetic fields, i.~e. their optical cross section is far greater than their geometrical cross section, and the associated strong local field enhancement in the proximity of the particle surface as depicted in Figs.~\ref{fig1}(c) and (e). Note that we have used a Drude model for gold ($\omega_{p,Au}=9\ $eV, $\epsilon_{b,Au}=9$, $\gamma_{p,Au}=0.071\ $eV), excluding intraband effects, for clarity.
This is put in comparison to the considered electrolyte system in Figs.~\ref{fig1}(d) and (f). The aforementioned effective LSPR for the coupled system agrees with the extinction cross section maximum and given the small membrane size of $R=10\ $nm considered here the light trapping is several orders of magnitude higher than for the gold particle of the same size. However, the associated local field enhancement at resonance is much smaller than for the metal. This can be understood and potentially remedied as follows. We have considered (neutral) water as background everywhere, while in the metal system a strong refractive index contrast is found at the surface. Due to the commonly known boundary condition of continuous $\epsilon \vec{\hat{n}}\cdot\vec E$ this yields an increase in the field on the outside of the spherical nanoparticle. In case of the ionic system, no such strong contrast is found, however, the lipid membrane used to confine the electrolyte could be filled with a different medium than water, e.~g. a cytoplasm. It should also be noted that the field enhancement increases for even lower energies until it saturates. This region is most strongly affected by non-classical charge carrier interactions.

Finally, we compare classical electrodynamic results (standard Mie coefficients) with our nonlocal model. For the metal system, as reported previously, a blueshift of the LSPR is observed and on inclusion of electron diffusion effects the resonance becomes more strongly broadened and attenuated, see Fig.~\ref{fig1}(e). The coupled ionic system also shows an attenuation, see Fig.~\ref{fig1}(f), which we study in depth in this article. The LSPR position is marked by a vertical line.

\begin{figure}
\centering
	\includegraphics[width=\linewidth]{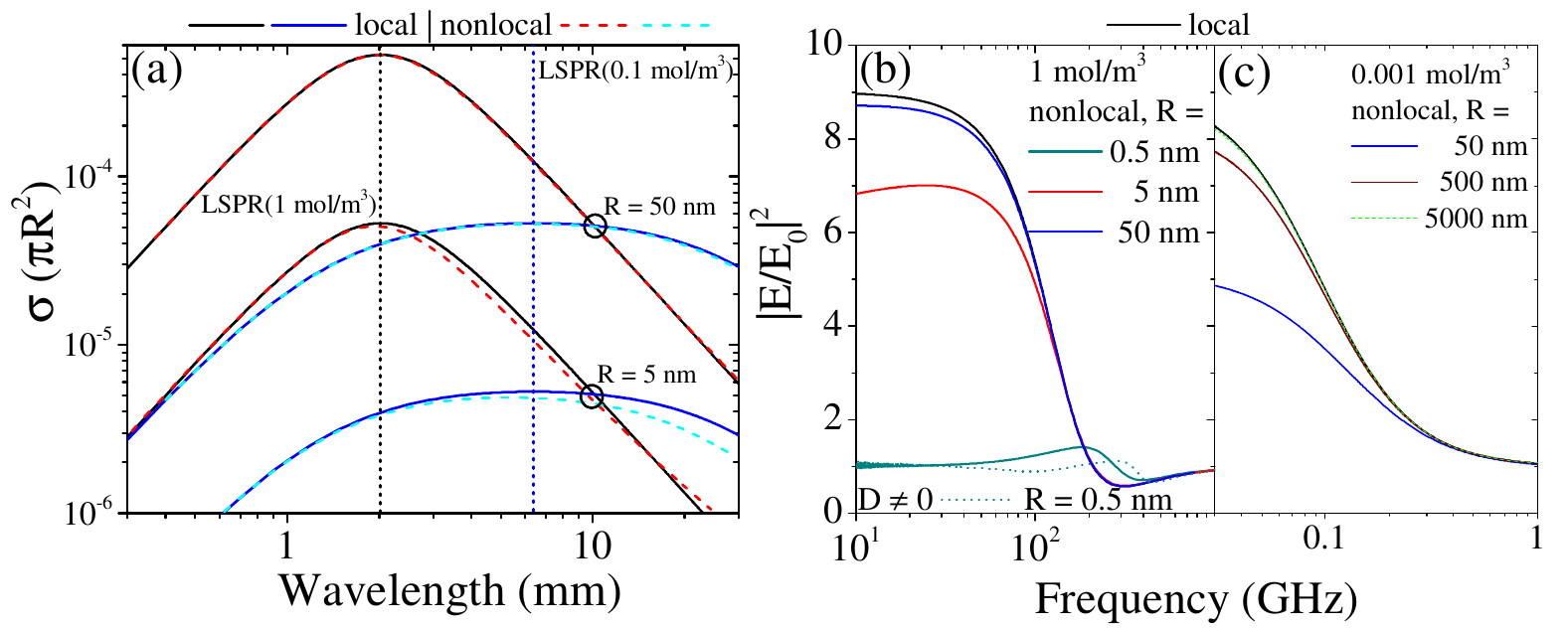}
    \caption{\textbf{Extinction cross section and quenching of field enhancement at different ion concentration levels and for several orders of magnitude in particle size.} (a) Extinction cross section normalized to geometrical cross section of the membrane. (b), (c) Comparing the maximum intensity at the particle surface for ionic systems with an ion concentration of (b) $n=1\nicefrac{\rm{mol}}{\rm{m}^3}N_A$ and (c) $n=10^{-3}\nicefrac{\rm{mol}}{\rm{m}^3}N_A$ for varying membrane size.}
    \label{fig2}
\end{figure}

First, we try to observe a blueshift of the LSPR with respect to the classical picture of the electrolyte in Fig.~\ref{fig2}(a). We consider two different ion concentrations that yield different resonance wavelengths and consider two membrane sizes, one that behaves classically ($R=50\ $nm) and one that shows nonlocal response ($R=5\ $nm). Zooming into the resonances (note the log-log axis), we indeed note a blueshift, though it is not as dramatic as for the metal system, since the resonances are much broader.

Hence, we concentrate on the quenching of the field enhancement and study its dependence on particle size for two different ion concentrations in Figs.~\ref{fig2}(b) and (c). The ion concentration determines the LSPR and frequency range of interest for the system. With this, the impact of nonlocal quenching of the local field can be shifted to larger particle sizes. In general, a large enough membrane size yields the local limit where nonlocal charge carrier dynamics are no longer sizable. We have once more included in Fig.~\ref{fig2}(b) for an extremely small membrane size, where the blueshift is remarkable. It should be noted, however, that a water molecule has a diameter of about 0.275\ nm. A membrane with radius $R=1\ $nm ($R=10\ $nm) could accommodate approximately 380 ($3.8\times10^5$) (densely packed) water molecules. Hence, it should be made clear, that the results presented for subnaometer radii have to be taken with care.

\begin{figure}
\centering
	\includegraphics[width=0.5\linewidth]{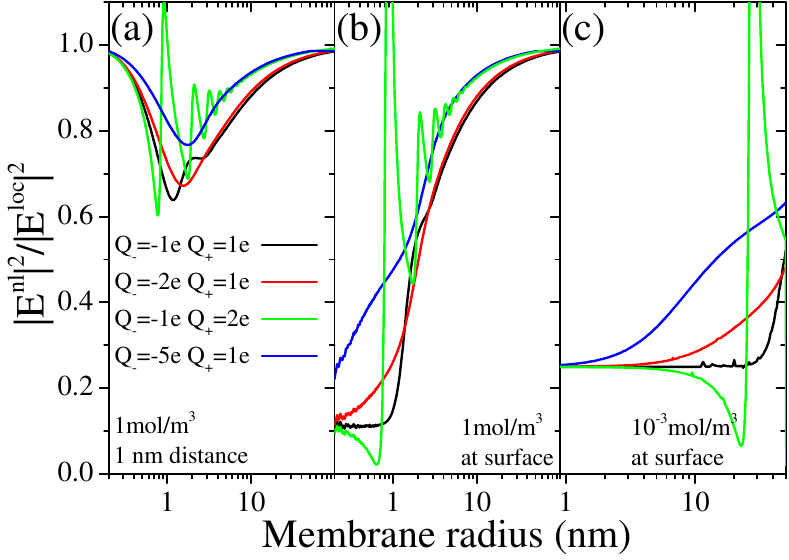}
    \caption{\textbf{Quenching of field enhancement as a function of membrane size and charge imbalance.} Comparing the ratio of the maximum intensity between local and nonlocal theories at (a) a 1\ nm distance and (b), (c) the membrane surface for ionic systems with an ion concentration of (a), (b)  $n_-=1\nicefrac{\rm{mol}}{\rm{m}^3}N_A$ evaluated at $\omega=10\ $GHz and (c) $n_-=10^{-3}\nicefrac{\rm{mol}}{\rm{m}^3}N_A$ evaluated at $\omega=10^{-2}\ $GHz as a function of membrane size for varying charge imbalance. Due to total charge balance $n_+Q_+=n_-|Q_-|$.}
    \label{fig3}
\end{figure}

We digress from the hydroxide and hydronium system in Fig.~\ref{fig3}. Here, we consider the local field quenching for membranes up to 200\ nm in diameter for an imbalance of charges in the ions at a fixed frequency, where the field enhancement is saturated for the $|Q_\pm|=e$. This yields an imbalance of ion concentration for the two types, balancing the total charge. Due to the different masses, it makes a difference which type of ions has the larger charge. Fig.~\ref{fig3}(a) shows that even at a distance of 1\ nm away from the membrane surface, a field quenching of up to 30\% can be found. Again, results for membrane sizes below $R=1\ $nm are purely academical, but the solution to the coupled equations of charge dynamics and electromagnetic wave equation are well defined. Figs.~\ref{fig3}(b) and (c) are evaluated directly at the membrane surface for different ion concentrations and even more dramatic nonlocal field quenching is observed, where the system with the largest charge imbalance shows the lowest quenching. This can be explained by noting that the LSPR of the systems is pushed further away from the excitation frequency.

\begin{figure}
\centering
	\includegraphics[width=0.5\linewidth]{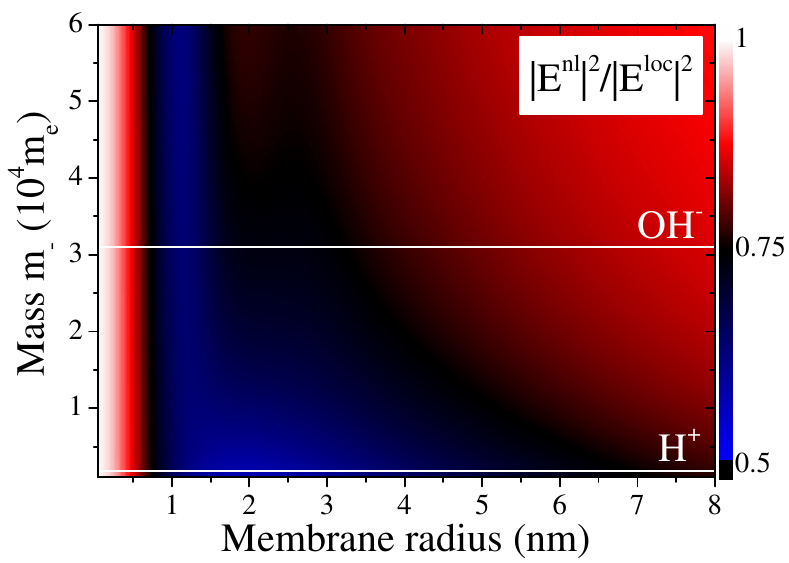}
    \caption{\textbf{Quenching of field enhancement as a function of membrane size and mass imbalance.} Comparing the ratio of the maximum intensity between local and nonlocal theories at a 1\ nm distance from the membrane surface for ionic systems with an ion concentration of $n_-=1\nicefrac{\rm{mol}}{\rm{m}^3}N_A$ evaluated at $\omega=10\ $GHz as a function of membrane size and mass. The white lines mark the masses of H$^+$ and the parametrized OH$^-$. White areas show classical behavior, about 10\% to 20\% quenching is given in red ares and black and blue areas are dominated by nonlocal charge dynamics.}
    \label{fig4} 
\end{figure}

A further imbalance of interest is the mass imbalance. We study the local field quenching due to spatial dispersion in Fig.~\ref{fig4}. Larger membrane radii yield a steady 20\% quenching until the nonlocal effects cease and both theories coincide. Tuning thus the optical properties of the ionic system, the nonlocal response can be made sizable at larger membrane sizes as apparent from Fig.~\ref{fig3}(c). For heavy ions paired with light ions (lower part of the graph) stronger quenching can be expected for a larger range of membrane sizes. For heavy ions paired with even heavier ions this range of strong attenuation is considerably smaller.

In the next subsections, we summarize to the main aspects of the developed theory which is further detailed in the Methods section.

\subsection*{Two-fluid model}
\begin{figure}
\centering
	\includegraphics[width=0.5\linewidth]{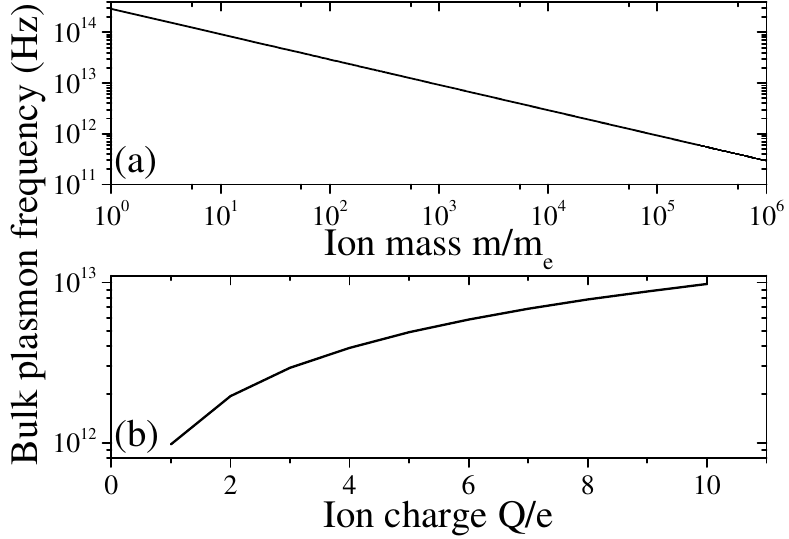}
    \caption{\textbf{Ionic bulk plasmon frequency values $\omega_p^2=\nicefrac{4\pi Q^2 n}{m}$.} (a) For different choices of the mass. (b) For different values of the ion charge. Fixed values are $Q=3e, m=10^4m_e, \epsilon_b=1.77, n=10^{-2}N_0$, where $N_0$ is the one-molar concentration.}
    \label{fig5}
\end{figure}

The optical response of an ionic system with free negative and positive charges in a fluid is based in the hydrodynamic model on separating the dynamics of both types of ions, assuming $\nabla\vec D = \nabla\epsilon_b\vec E = 4\pi(\rho_-+\rho_+)$. Hereby, $\vec D$ is the displacement vector, $\vec E$ the electric field vector, $\epsilon_b$ the dielectric background permittivity of the solution and $\rho_\pm$ are the (external) charge densities. Ion masses $m_\pm$ and charge $Q_\pm$ may differ for the ions, however, the total charge is balanced, i.~e. the charges are mutually compensated by setting $|n_+Q_+|\equiv |n_-Q_-|$. This yields an equal density $n=n_+=n_-$ of charges in our case of hydronium and hydroxide with equal, but opposite charges.
The electromagnetic wave equation then reads
\begin{align}\label{eq.WEQstart}
 \nabla\times\nabla\times\vec E-k^2\epsilon_b\vec E = \frac{4\pi i k^2}{\omega}\left( \vec j_-+\vec j_+\right).
\end{align}
Hereby, $\vec j_\pm$ are the charge current densities of the ions in the electrolyte. The electromagnetic properties are evaluated at frequency $\omega=ck$ connected to the wave vector $k$ via the speed of light $c$. Together with the hydrodynamic equations for each type of ion, see Methods section for details, this can be brought into the form~\cite{David2011a} of
\begin{align}\label{eq.WEQ}
  \nabla^2\vec E + k^2\epsilon_\perp^{\rm{ions}}\vec E = 4\pi\nabla(\eta_-\rho_-+\eta_+\rho_+),
\intertext{where the right hand side vanishes in the local limit, $\eta_\pm$ comprise of material dependent parameters and we have defined the permittivity of the coupled ion system as}
\epsilon_\perp^{\rm{ions}} = \epsilon_b - \frac{\omega_{p-}^2}{\omega(\omega+i\gamma_-)} - \frac{\omega_{p+}^2}{\omega(\omega+i\gamma_+)}.
\end{align}
Note that we obtain the common expressions for the (ionic) bulk plasmon frequencies $\omega_{p\pm}^2=\nicefrac{4\pi Q_\pm^2 n_\pm}{m_\pm}$ without introducing an auxiliary jellium in which the ions are contained. 
The range of the ionic bulk plasmon frequency is investigated in Fig.~\ref{fig5} with respect to the ion mass and charge which can be selected via the material. The ionic plasmon frequency is far smaller than that of common metals and lies in the infrared and beyond ($>3\mu$m).

However, in order to study specific geometries of the resulting plasmon oscillations, the electrolyte system has to be confined in the desired way. This can be achieved experimentally via impermeable membranes. Fig.~\ref{fig1} shows the similarities between the plasmons in a solid metal nanoparticle (gold in a simplified Drude model) and the electrolyte system at increased density in terms of both the extinction cross section and field enhancement.

\subsection*{Navier-Stokes equation}
In the hydrodynamic model, we determine the induced current density $\vec j_\pm$ with the linearized Navier-Stokes equation for a charged plasma 
\begin{align}
	\vec j_\pm=\frac{i}{\omega+i\gamma_\pm}\left(\frac{Q_\pm^2 n_\pm}{m_\pm}\vec E - \nabla\beta_{\rm{GNOR}\pm}^2\rho_\pm\right). \label{eq.hydro}
\end{align}
The pressure term is derived from classical gas theory of a Thomas-Fermi gas~\cite{Huang1963,David2011a,David2014a} as $\frac{p}{m}=\frac{1}{3}v_F^2n=\beta^2n$ which we adopt for our present system. Including further quantum mechanical effects, in particular Coulomb interaction, it was shown that the nonlocal interaction strength~\cite{Pines1952} is rather $\beta^2\equiv\frac{3}{5}v_F^2$ for metals, where $v_F$ is the Fermi velocity of the material defined as $\hbar \vec k_F=\vec p = \vec v_F m_e.$
In case of heavy ions, we rely on the thermal velocity of the ions $v_{\rm{th}}=\sqrt{\nicefrac{3k_B T}{m}}$, which is much smaller than the velocities defined for free charges in solid materials ($v_F=1.4\times 10^6$m/s for Au and Ag) due to the large masses involved ($v_{\rm{th}}\sim10^3$m/s for hydroxide and hydronium, see table~\ref{tab.mat}).

Note that for positive charges $\beta_+=-\beta_-$, due to integration over the valence band below the Fermi energy for positive charges, rather than over the conduction band above the Fermi energy for negative charges.

Finally, the generalized nonlocal optical response (GNOR~\cite{Mortensen2014,Raza2015a}) allows a straightforward extension to also include diffusion effects via a diffusion constant $D$. This is obtained setting $\vec j \rightarrow \vec j - D\nabla\rho$ and results in an effectively complex-valued and wavelength-dependent $\beta_{\rm{GNOR}}^2=\beta^2+D(\omega+i\gamma)$. The local limit is obtained for $\beta_{\rm{GNOR}}\rightarrow0$ which suppresses the interaction of the charged particles in the fluid. However, the field induced current density remains and yields the ionic plasmon effects within classical electrodynamics.

For solid particles, nonlocal interaction results in plasmon broadening and quenching~\cite{}. This is also observed for the electrolyte as depicted in Figs.~\ref{fig1} to \ref{fig4}. The typically observed blueshift of the plasmon resonance with respect to the local result is also found in the electrolyte system, see Fig.~\ref{fig2}(a), but due to much larger wavelengths does not critically alter the classical resonance position.

The minimum value for the diffusion constant to show an additional effect was found to be $D_\pm=0.5\times10^{-5}$m$^2$/s (with $\gamma_\pm\approx10^{11}$ Hz, hence $D_\pm\gamma_\pm\approx5\times 10^{5}$m$^2$/s), where in comparison $|\beta_\pm|\approx1.5\times10^5$(m/s)$^2$.

\subsection*{Coupled nonlocal wave equations}
From the continuity equation $i\omega\rho_\pm = \nabla\vec j_\pm$ we find for each type of charge a wave equation. Introducing the transversal permittivity for the respective charges as $\epsilon_{\perp\pm}=\epsilon_b-\frac{\omega_{p\pm}}{\omega(\omega+i\gamma_\pm)}$ and using $\nabla E=\nicefrac{4\pi(\rho_-+\rho_+)}{\epsilon_b}$, we obtain coupled hydrodynamic equations typical for a two-fluid model
\begin{align}
	\rho_-= \frac{\epsilon_b}{\omega_{p+}^2}\left(\nabla^2\beta_{\rm{GNOR+}}^2+\frac{\omega(\omega+i\gamma_+)}{\epsilon_b}\epsilon_{\perp+} \right)\rho_+
\end{align}
and vice versa. The analytic solution yields a wave equation of fourth order
\begin{align}
	0 &=\beta_{\rm{GNOR+}}^2\beta_{\rm{GNOR-}}^2\left( \nabla^2 + q^2_+ \right) \left( \nabla^2 + q^2_-\right)\rho_--\frac{\omega_{p+}^2\omega_{p-}^2}{\epsilon_b^2}\rho_-, \label{eq.nlWEQ}
\end{align}
which can be solved analytically. Note that we introduced the nonlocal wave vectors $q_\pm$ from the single charged plasma result~\cite{David2011a}
$q_\pm^2=\frac{1}{\beta_{\rm{GNOR\pm}}^2}\frac{\omega(\omega+i\gamma_\pm)}{\epsilon_b}\epsilon_{\perp\pm}$.

\begin{figure}
\centering
  \includegraphics[width=\columnwidth]{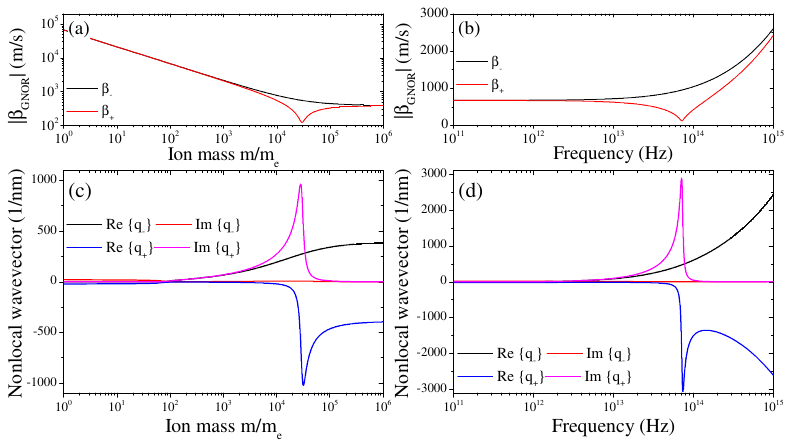}
  \caption{\textbf{Nonlocal properties of the ions.} (a), (b) The nonlocal strengths $\left|\beta_{\rm{GNOR\pm}}\right|$ and (c), (d) the nonlocal wavevectors $q_\pm$ for positive and negative charges are studied as a function of the (a), (c) ion mass and (b), (d) photon energy.}
  \label{fig6}
\end{figure}

We show in Fig.~\ref{fig6} the values of $\left|\beta_{\rm{GNOR\pm}}\right|$ and $q_\pm$ as functions of the ion masses and the frequency of incoming light. Due to the change in sign for the positive charges, the nonlocal coupling strength can be compensated by the diffusion term. This is the local limit $\beta_{\rm{GNOR\pm}}\longrightarrow0\,\Rightarrow\, q_\pm\longrightarrow\infty$, wich, however, does not change the (local) plasmonic character of the ionic system under irradiation.

Both the imaginary parts are positive by design, but the real part, responsible for the direction of the oscillation differs by both its sign and strength as could be expected from the analytic expressions. The positive charges have thus the opposite oscillation direction and an inherent resonance structure that makes the nonlocal properties vanish faster than what is found for negative charges.

\subsection*{Finite spherical ionic systems}
The practical biological cell organization can be thought of like this: A lipid cellular membrane contains a  liquid or a cytoplasm with the ions embedded in it. The ionic system is thus trapped into a specific geometry and we investigate the model further to describe the optical response of such geometries. This has many parallels with the theory of a single electron plasma in metal nanoparticles~\cite{David2011a} for which we find Mie parameters extended by a single nonlocal term. This exploits the usual boundary conditions and additional boundary conditions stemming from the nonlocal wave equations for each type of ion, see Methods section for details. 

The scattering matrix for the two-fluid, spherical ionic system becomes in full, formal analogy
\begin{align}
	t_l^{E}&=\frac{-\epsilon_\perp j_l(\theta_\perp)[\theta_0j_l(\theta_0)]' +\epsilon_bj_l(\theta_0)([\theta_\perp j_l(\theta_\perp)]'+g_{l}^{\rm{ions}})}{\epsilon_\perp j_l(\theta_\perp)[\theta_0h_l^+(\theta_0)]' -\epsilon_bh_l^+(\theta_0)([\theta_\perp j_l(\theta_\perp)]'+g_{l}^{\rm{ions}})},
	\label{eq.ionic_tE}
\end{align}
with $g_l^{\rm{ions}}=g_{l+}+g_{l-}$
\begin{align}
g_{l\pm}&=\frac{l(l+1)j_l(\Theta^{\rm{ions}})}{R}\frac{j_l(q_\pm R)}{q^{\rm{Mie}}_\pm j'(q^{\rm{Mie}}_\pm R)} \left( \frac{\epsilon_{\perp\pm}}{\epsilon_b}-1 \right).
\end{align}
Note that the nonlocal parameter $g_l^{\rm{ions}}$ vanishes under the assumption of local response fully recovering the original Mie coefficients~\cite{Mie1908, Jackson1976}. This allows us to study the electro-optical properties of spherical membranes filled with a nonlocal electrolyte with only a small correction in available numerical procedures. This is very similar to the single plasma result, where we now sum nonlocal contributions from two charge plasmas, see Methods section for a comparison.

The main differences lie in the definition of $q^{\rm{Mie}}$ and the fact, that we have two contributions from different type of charge carriers. Hence, an ionic system where the charge densities, masses and charges can possibly be different may lead to intriguing results more tunable than . The main difference is also the sign in $\beta_{\rm{GNOR\pm}}$. Significantly different amplitudes can be expected when the masses and charges differ, e.~g. $m_-\ll m_+$.

\section*{Discussion}
The presented framework on nonlocal spherical ionic systems is bridging hard and soft matter theory by offering insights into non-classical effects in electrolytes in terms of plasmonic properties of oscillating charge carriers. We have relied on classical gas theory, where ionic charge carriers move much slower than free electrons in metals, and low diffusion parameters. In addition, in the absence of a solid crystal lattice as in metal nanoparticles, the probability of scattering events becomes strongly reduced, which is reflected in the small plasmon damping coefficients for the considered electrolyte system, see table~\ref{tab.mat}. Despite these highly classical parameters resulting in a low nonlocal coupling strength $\beta$, we found that nonlocal charge interaction plays a sizable role in ionic systems mediated by the coupling of their dynamics and interaction with an external electromagnetic field. In contrast to spatial dispersion in solid metal particles, this is found even at large particle sizes and low plasmon frequencies. These nonlocal effects arise from both positive and negative charge dynamics and show an impact in the (far-) infrared depending on the composition of the ionic system being tunable via their charges and masses.

In this article, we have studied hydroxide and hydronium, which have a charge of $\pm1e$ and compared to other electrolyte systems low overall mass $\approx 3\times10^5m_e$. The plasmon frequency of an ionic system is increased drastically with the charge ($\omega_p\sim Q$) and reduced moderately with the mass ($\omega_p^2\sim1/m$). Hence, the usage of other electrolytes with view to (photo-) catalysis allows tuning the LSPR frequency over a broad infrared range, see also Fig.~\ref{fig6} in the Methods section.

Nonlocal soft plasmonics is considered highly relevant for biological systems, e.~g. particle chains representing aligned axons. The inclusion of further non-classical effects such as Lorentz friction~\cite{Jacak2015} into the theory developed here is straightforward thanks to analytic expressions in terms of modified damping terms $\gamma$. This has been considered successfully for metal nanoparticles in Ref.~\cite{Kluczyk2017}
and awaits further study in the scope of ionic systems.

With view to photocatalytic applications, planar structures are of further interest to investigate plasmonic effects of charged ions. For electrolytes confined in a planar system, the local limit is additionally given for light at normal incidence. The vanishing parallel momentum of the incoming light suppresses coupling to longitudinal modes in the ionic system. This will be discussed elsewhere.

\section*{Methods}
We use Maxwell's equations in Gauss units and neglect magnetic material properties, i.~e. $\mu=1$.

\begin{figure}
\centering
  \includegraphics[width=0.5\columnwidth]{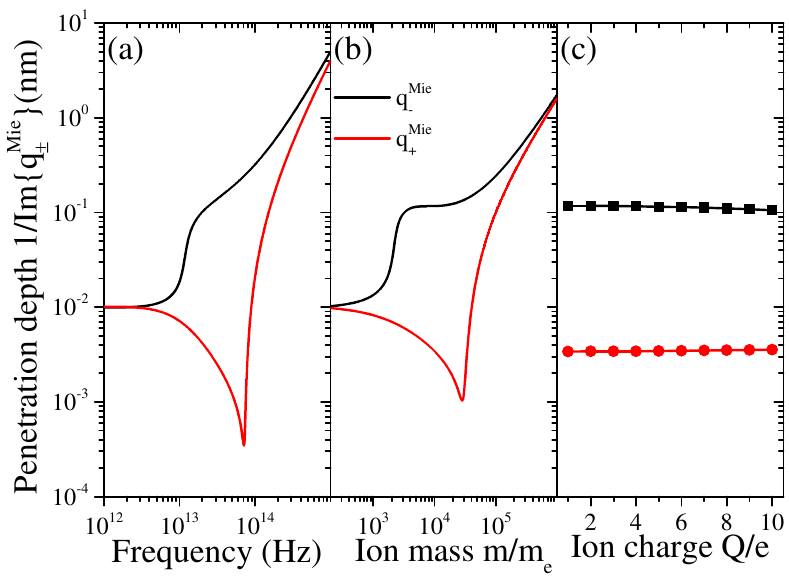}
  \caption{\textbf{Penetration depth of nonlocal pressure waves for the coupled ionic system with nonlocal properties.} We compare the inverse of the imaginary part of exact solutions as a function of (a) photon energy, (b) ion mass and (c) ion charge.}
  \label{fig7}
\end{figure}

We determine the induced current density $\vec j$ with the Navier-Stokes equation for a charged plasma
\begin{align}
	-Qn\left( \partial_t + \vec v\nabla\right) \vec v -\gamma Qn\vec v= \frac{Q^2 n}{m}\left( \vec E + \vec v\times\vec B\right) + \frac{1}{m}\nabla p.
\end{align}
Note that $-Qn=\rho$ and $-Qn\vec v=\rho\vec v=\vec j$, so that $Q=e$ for electrons and $Q=-e$ for positrons. We aim to linearize the Navier-Stokes equation with respect to the charge density. We can therefore set $B\equiv0$ and $\vec v\nabla\equiv0$. Furthermore, we assume all quantities to be first order in their response $n = n_0 + \delta n$, such that $\vec j \equiv -Qn_0\vec v$ and $p=p_0+\delta p$ and $\vec E = \vec E_0+\delta E$. We obtain a zeroth order equation defining how the static pressure $p_0$ is compensated by a static field $E_0$. The remaining first order expression (where we neglect the symbol for variation $\delta$ for readability) results in eq.~\eqref{eq.hydro}. We insert this into eq.~\ref{eq.WEQstart} and obtain
\begin{align}
	\nabla\times\nabla\times\vec E - k^2\epsilon_\perp^{\rm{ions}}\vec E &= 4\pi\frac{k^2}{\omega}\nabla\left(\frac{\beta_{\rm{GNOR-}}^2}{\omega+i\gamma_-}\rho_- + \frac{\beta_{\rm{GNOR+}}^2}{\omega+i\gamma_+}\rho_+\right),\\
    \nabla^2\vec E + k^2\epsilon_\perp^{\rm{ions}}\vec E &= 4\pi\nabla \left(\left(\frac{1}{\epsilon_b}-\frac{k^2\beta_{\rm{GNOR-}}^2}{\omega(\omega+i\gamma_-)}\right)\rho_- + \left(\frac{1}{\epsilon_b} - \frac{k^2\beta_{\rm{GNOR+}}^2}{\omega(\omega+i\gamma_+)}\rho_+\right)\right),
\end{align}
which defines $\eta_\pm=\frac{1}{\epsilon_b} - \frac{k^2\beta_{\rm{GNOR+}}^2}{\omega(\omega+i\gamma_+)}$.

\subsection*{Spherical membrane}
We derive the Mie coefficients~\cite{Mie1908} for a spherical membrane of radius $R$ filled with an electron plasma and positively charged ions subject to an electromagnetic field using the two-fluid model.

It is convenient to use an expansion of the electric field into scalar functions~\cite{Low1997} as 
\begin{align}
	\vec E=(1/k)\nabla\psi^{L}+\vec L\psi^{M}+\frac{\nabla\times\vec L}{ki}\psi^{E},
\end{align}
where $\vec L=-i\vec r\times\nabla$ is the angular momentum operator, and the superscripts $E$, $M$, and $L$ indicate electric, magnetic, and longitudinal components, respectively. The additional boundary condition in spherical coordinates $\vec \hat{e}_r\vec j=0$ yields
\begin{align}
      \beta_{\rm{GNOR}\pm}^2\deriv{r}{} \rho_\pm&= \frac{Q_\pm^2 n_{\pm}}{m_\pm k}\left(\deriv{r}{}\psi^{L} + \frac{1}{r}l(l+1)\psi^{E} \right) \label{eq.ABC}
\end{align}
in terms of the scalar functions and the angular momentum number $l$ using the identity $-\vec r\cdot(\nabla\times\vec L)/i=L^2=l(l+1)$. The common boundary conditions for the electric and magnetic field components result in the continuity of $\psi^{M}$, $(1+r\deriv{r}{})\psi^{M}$, $\psi^{L}+(1+r\deriv{r}{})\psi^{E}$, and $\epsilon\psi^{E}$ for the scalar functions. The magnetic and electric scalar functions $\psi^{\nu}\,(\nu= \{E,M\})$ obey a Helmholtz equation of the form $(\nabla^2+k^2\epsilon^{\rm{ions}})\psi^{\nu}=0$ and can therefore be expanded in terms of spherical Bessel functions $\psi^{\nu}=\sum_L\psi^{\nu}_Lj_L(k^{\rm{ions}} r)$. Similarly, the electron density is expanded into $\rho^{\rm{ind}}(\vec r,\omega) = \sum_L(\rho_{l+}j_L(q^{\rm{Mie}}_+r) + \rho_{l-}j_L(q^{\rm{Mie}}_-r))$.

The longitudinal wave vectors $q^{\rm{Mie}}_\pm$ given by $(q^{\rm{Mie}_\pm})^2=-\frac{\tilde{p}}{2}\pm\sqrt{\left(\frac{\tilde{p}}{2}\right)^2-\tilde{q}}$ are the solutions to eq.~\eqref{eq.nlWEQ} with
\begin{align}
 	\tilde{p}=- \left(q_+^2 + q_-^2\right),\qquad
    \tilde{q}=-\left(\frac{1}{\beta_{\rm{GNOR-}}^2\beta_{\rm{GNOR+}}^2}\frac{\omega_{p+}^2\omega_{p-}^2}{\epsilon_b^2} - q_-^2q_+^2\right).
\end{align}
We demand $\Im\{q^{\rm{Mie}}_\pm\}>0$, i.~e. that nonlocal, longitudinal excitations remain absorptive, which reduces the four solutions to two relevant ones. 
We study the wave vector of the combined ion system $q^{\rm{Mie}}_\pm$ in terms of the penetration depths $\nicefrac{1}{|\Im\{q_\pm\}|}$ in Fig.~\ref{fig7} for a range of (a) photon energies of the incoming light, (b) ion masses and (c) charges. The latter shows almost no effect, but the former two a yield sizable nonlocal skin effect. Hereby, the wavevector associated to the positive charge carriers can be suppressed in the same wave as $\beta_{\rm{GNOR+}}$. 

The longitudinal scalar function satisfies a different wave equation, namely $\nabla^2\psi^L=4\pi k(\rho_{l-} +\rho_{l+})/\epsilon_b$, which we find from the Coulomb law $\nabla\epsilon_b\vec E=4\pi(\rho_-+\rho_+)$. From this and the wave equation eq.~\eqref{eq.WEQ}, we deduce
\begin{align}
 	(\nabla^2+k^2\epsilon^{\rm{ions}})\psi^L=4\pi k\left(\eta_-\rho_{l-} +\eta_+\rho_{l+}\right),\quad
    \Rightarrow\quad\psi^L=-\frac{4\pi k}{\epsilon^{\rm{ions}}}\left(\frac{\beta_{\rm{GNOR-}}^2}{\omega(\omega+i\gamma_-)}\rho_{l-} + \frac{\beta_{\rm{GNOR+}}^2}{\omega(\omega+i\gamma_+)}\rho_{l+}\right).
\end{align}

Note that the above analysis is valid for the spherical region that contains the ionic system, where the electric ($\nu=E$) and magnetic ($\nu=M$) field are given by $A^\nu_lj_L$, with $j_L=j_{lm}(k_\perp r)$. Outside the particle-like region, the longitudinal scalar function vanishes since there are no induced charges in the dielectric surrounding. Therefore, the scalar field is given by $j_{lm}(k_0 r)+t_l^\nu h^+_{lm}(k_0 r)$ around the particle. Unknown parameters are the amplitude $A^\nu_l$ inside and scattering matrix $t_l^\nu$ outside. Exploiting the boundary conditions stated above, we find a set of linear equations for the magnetic and electric  scattering matrices. Interestingly, the magnetic scattering matrix is unchanged with respect to the local theory, indicating that magnetic modes are not sensitive to the induced longitudinal modes by neither specimen. The scattering matrix for the electric scalar function is more complicated than in the local approximation due to the appearance of $\psi^{L}$ in the charged region that contains information on the nonlocal response. At this point, the only difference to previous results for nonlocal electron dynamics in metals nanoparticles, where only free electrons are considered, is the more involved expression for the wave vectors.

We want to solve the following set of equations for the electric scalar field.
\begin{align}\label{eq.BC1}
    \epsilon^{\rm{ions}}A_l^Ej_l(\Theta^{\rm{ions}}) = \epsilon_0\left(j_l(\Theta_0)+t_l^Eh_l^+(\Theta_0)\right),\\\label{eq.BC2}
-\frac{4\pi k}{\epsilon^{\rm{ions}}}\left(\frac{\beta_{\rm{GNOR-}}^2\rho_{l-}j'(q^{\rm{Mie}}_-R)}{\omega(\omega+i\gamma_-)} + \frac{\beta_{\rm{GNOR+}}^2\rho_{l+}j'(q^{\rm{Mie}}_+R)}{\omega(\omega+i\gamma_+)}\right)+A_l^E\left[\Theta^{\rm{ions}}j_l(\Theta^{\rm{ions}})\right]'\nonumber\\= \left[\Theta_0j_l(\Theta_0)\right]'+t_l^E\left[\Theta_0h_l^+(\Theta_0)\right]',
\end{align}
where $\Theta_\mu=k_\mu R\sqrt{\epsilon_\mu}$.
We use the additional boundary conditions, eq.~\ref{eq.ABC}, for the charged plasmas to find expressions allowing the substitution of $\rho_{l\pm}$ in the above equations.
\begin{align}
	\beta_{\rm{GNOR-}}^2\rho_{l-}q^{\rm{Mie}}_-&j'(q^{\rm{Mie}}_-R) = \frac{\omega_{p-}^2l(l+1)}{4\pi kR}A_l^Ej_l(\Theta^{\rm{ions}}) \nonumber\\
    &-\frac{\omega_{p-}^2}{\epsilon^{\rm{ions}}}\left(\frac{\beta_{\rm{GNOR-}}^2q^{\rm{Mie}}_-j'(q^{\rm{Mie}}_-R)}{\omega(\omega+i\gamma_-)}\rho_{l-} + \frac{\beta_{\rm{GNOR+}}^2q^{\rm{Mie}}_+j'(q^{\rm{Mie}}_+R)}{\omega(\omega+i\gamma_+)}\rho_{l+}\right)
\end{align}
and vice versa. From this, we find for the charges
\begin{align}
	\rho_{l\pm}&=A_l^E\frac{l(l+1)}{4\pi kR}\frac{j_l(\Theta^{\rm{ions}})}{q^{\rm{Mie}}_\pm j'(q^{\rm{Mie}}_\pm R)}\frac{\epsilon^{\rm{ions}}}{\epsilon_b}\frac{\omega_{p\pm}^2}{\beta_{\rm{GNOR\pm}}^2}.
\end{align}
Next, we insert the charge densities into the equations obtained with the common boundary conditions \eqref{eq.BC1} and \eqref{eq.BC2} and rewrite after some algebra
\begin{align*}
    A_l^E\left\{\left[\Theta^{\rm{ions}}j_l(\Theta^{\rm{ions}})\right]' + g_l^{\rm{ions}}\right\}&= \left[\Theta_0j_l(\Theta_0)\right]'+t_l^E\left[\Theta_0h_l^+(\Theta_0)\right]',
\end{align*}

\subsection*{Comparing to an electron plasma in solid metal}

The nonlocal contribution to the field via the induced charge density for a single electron plasma results in $\rho=A_l^E\frac{l(l+1)}{4\pi kR}\frac{j_l(\Theta_\perp)}{qj'(qR)}\frac{\epsilon_\perp}{\epsilon_b}\frac{\omega_{p}^2}{\beta^2}$ for metals~\cite{David2011a}. The local scattering matrix can then be extended by a single parameter describing nonlocal behavior of the electron motion in the conduction band
\begin{align}
	g_l&= \frac{l(l+1)j_l(\theta_\perp)j_l(qa)}{qaj_l'(qa)} \left( \frac{\epsilon_\perp}{\epsilon_b}-1 \right)
    \label{eq.gl}
\end{align}
and becomes with $\theta_0=ka\sqrt{\epsilon_0}$ and $\theta_\perp=ka\sqrt{\epsilon_\perp}$.
\begin{align}
	t_l^{E}&=\frac{-\epsilon_\perp j_l(\theta_\perp)[\theta_0j_l(\theta_0)]' +\epsilon_0j_l(\theta_0)([\theta_\perp j_l(\theta_\perp)]'+g_l)}{\epsilon_\perp j_l(\theta_\perp)[\theta_0h_l^+(\theta_0)]' -\epsilon_0h_l^+(\theta_0)([\theta_\perp j_l(\theta_\perp)]'+g_l)},
	\label{eq.hydroMIE_tE}
\end{align}
where the primes indicate differentiation with respect to the $\theta$ variables.

\section*{Data Availability}
The datasets generated and analyzed during the current study are available from the corresponding author on reasonable request.


\section*{Additional information} \paragraph{Competing interests} The author declares no competing interests.

\section*{Acknowledgements}
The author thanks L. Jacak and W.~A. Jacak for fruitful discussions. Financial support from the European AMAROUT-II program (Grant Agreement number 291803) and from the Spanish Ministry of Economy, Industry and Competition (MINECO) via funding of the Centers of Excellence Severo Ochoa (Ref. SEV-2016-0686) is acknowledged. Financial support is further provided by the Comunidad de Madrid (Ref. 2017-T2/IND-6092). This work was supported by the European Cooperation in Science and Technology (COST) Action MP1406 MultiscaleSolar.

\section*{Author contributions statement}
C.~D. developed the theory, produced and analyzed the presented numerical results. 

\begin{table}\centering
\begin{tabular}{|c|c|c|c|c|c|}\hline
Ion 		& mass (Da) & mass ($m_e$) & thermal velocity $v_{\rm{th}}$ (m/s) & free mean path (nm) & damping (meV)\\\hline
H$^+$		& 1.007		 &  1835		 & 2727	& 0.33	& 17.22\\
OH$^-$		& 17.008	 & 31005		 &  663	& 1.35	&  1.02\\
H$_3$O$^+$	& 19.02		 & 34670		 &  627	& 1.42	&  0.91\\\hline
\end{tabular}
\caption{\bf{Electrolyte parameters for an aqueous solution in thermodynamic equilibrium.} The charges are $Q_\pm=\pm 1e$.}
\label{tab.mat}
\end{table}

\setcounter{figure}{0} 

\begin{figure}[ht]
\caption{\textbf{Concept of (localized surface) plasmon resonance (LSPR) in ionic systems compared to metals.} Illustration of charge carrier displacement in (a) a metal solid and (b) an ionic fluid confined by e.~g. a lipid cellular membrane. (c), (d) Field enhancement map of the plasmonic resonance in both systems. (e), (f) Impact of nonlocal charge dynamics (with and without electron diffusion $D$) on metals and ions in terms of the extinction cross section (upper panel) and the field enhancement (lower panel). The particle size is $R=10\ $nm and a Drude model for gold was used for clarity.}
\caption{\textbf{Extinction cross section and quenching of field enhancement at different ion concentration levels and for several orders of magnitude in particle size.} (a) Extinction cross section normalized to geometrical cross section of the membrane. (b), (c) Comparing the maximum intensity at the particle surface for ionic systems with an ion concentration of (b) $n=n_+=n_-=1\nicefrac{\rm{mol}}{\rm{m}^3}N_A$ and (c) $n=10^{-3}\nicefrac{\rm{mol}}{\rm{m}^3}N_A$ for varying membrane size.}
\caption{\textbf{Quenching of field enhancement as a function of membrane size and charge imbalance.} Comparing the ratio of the maximum intensity between local and nonlocal theories at (a) a 1\ nm distance and (b), (c) the membrane surface for ionic systems with an ion concentration of (a), (b)  $n_-=1\nicefrac{\rm{mol}}{\rm{m}^3}N_A$ evaluated at $\omega=10\ $GHz and (c) $n_-=10^{-3}\nicefrac{\rm{mol}}{\rm{m}^3}N_A$ evaluated at $\omega=10^{-2}\ $GHz as a function of membrane size for varying charge imbalance. Due to total charge balance $n_+Q_+=n_-|Q_|$.}
\caption{\textbf{Quenching of field enhancement as a function of membrane size and mass imbalance.} Comparing the ratio of the maximum intensity between local and nonlocal theories at a 1\ nm distance from the membrane surface for ionic systems with an ion concentration of $n_-=1\nicefrac{\rm{mol}}{\rm{m}^3}N_A$ evaluated at $\omega=10\ $GHz as a function of membrane size and mass. The white lines mark the masses of H$^+$ and the parametrized OH$^-$. White and red areas show classical behavior, while black and blue areas are dominated by nonlocal charge dynamics.}
\caption{\textbf{Ionic bulk plasmon frequency values $\omega_p^2=\nicefrac{4\pi Q^2 n}{m}$.} (a) For different choices of the mass. (b) For different values of the ion charge. Fixed values are $Q=3e, m=10^4m_e,\epsilon_b=1.77, n=10^{-2}N_0$, where $N_0$ is the one-molar concentration.}
\caption{\textbf{Nonlocal properties of the ions.} (a), (b) The nonlocal strengths $\left|\beta_{\rm{GNOR\pm}}\right|$ and (c), (d) the nonlocal wavevectors $q_\pm$ for positive and negative charges are studied as a function of the (a), (c) ion mass and (b), (d) photon energy.}
\caption{\textbf{Penetration depth of nonlocal pressure waves for the coupled ionic system with nonlocal properties.} We compare the inverse of the imaginary part of exact solutions as a function of (a) photon energy, (b) ion mass and (c) ion charge.}
\end{figure}

\end{document}